\input harvmac
\input amssym.def
\input amssym.tex
\noblackbox
\newif\ifdraft

\catcode`\@=11
\newif\iffrontpage
\newif\ifxxx
\xxxtrue

\newif\ifad
\adtrue
\adfalse

\newif\iffigureexists
\newif\ifepsfloaded
\def\epsfcheck{
\ifdraft
\input epsf\epsfloadedtrue
\else
  \openin 1 epsf
  \ifeof 1 \epsfloadedfalse \else \epsfloadedtrue \fi
  \closein 1
  \ifepsfloaded
    \input epsf
  \else
\immediate\write20{NO EPSF FILE --- FIGURES WILL BE IGNORED}
  \fi
\fi
\def\epsfcheck{}}
\def\checkex#1{
\ifdraft
\figureexistsfalse\immediate%
\write20{Draftmode: figure #1 not included}
\figureexiststrue 
\else\relax
    \ifepsfloaded \openin 1 #1
        \ifeof 1
           \figureexistsfalse
  \immediate\write20{FIGURE FILE #1 NOT FOUND}
        \else \figureexiststrue
        \fi \closein 1
    \else \figureexistsfalse
    \fi
\fi}
\def\missbox#1#2{$\vcenter{\hrule
\hbox{\vrule height#1\kern1.truein
\raise.5truein\hbox{#2} \kern1.truein \vrule} \hrule}$}
\def\lfig#1{
\let\labelflag=#1%
\def\numb@rone{#1}%
\ifx\labelflag\UnDeFiNeD%
{\xdef#1{\the\figno}%
\writedef{#1\leftbracket{\the\figno}}%
\global\advance\figno by1%
}\fi{\hyperref{}{figure}{{\numb@rone}}{Fig.{\numb@rone}}}}
\def\figinsert#1#2#3#4{
\epsfcheck\checkex{#4}%
\def\figsize{#3}%
\let\flag=#1\ifx\flag\UnDeFiNeD
{\xdef#1{\the\figno}%
\writedef{#1\leftbracket{\the\figno}}%
\global\advance\figno by1%
}\fi
\goodbreak\midinsert%
\iffigureexists
\centerline{\epsfysize\figsize\epsfbox{#4}}%
\else%
\vskip.05truein
  \ifepsfloaded
  \ifdraft
  \centerline{\missbox\figsize{Draftmode: #4 not included}}%
  \else
  \centerline{\missbox\figsize{#4 not found}}
  \fi
  \else
  \centerline{\missbox\figsize{epsf.tex not found}}
  \fi
\vskip.05truein
\fi%
{\smallskip%
\leftskip 4pc \rightskip 4pc%
\noindent\ninepoint\sl \baselineskip=11pt%
{\bf{\hyperdef\hypernoname{figure}{{#1}}{}}~}#2%
\smallskip}\bigskip\endinsert%
}


\def\{{\lbrace}
\def\}{\rbrace}

\def\a{\alpha}

\def\p{\partial}

\def\gz{\mathrel{\mathop g^{\scriptscriptstyle{(0)}}}} 
\def\gN{\mathrel{\mathop g^{\scriptscriptstyle{(N/2)}}}} 

\def\Box#1{\mathop{\mkern0.5\thinmuskip
           \vbox{\hrule\hbox{\vrule\hskip#1\vrule height#1 width 0pt\vrule}
           \hrule}\mkern0.5\thinmuskip}}
\def\Laplace{\displaystyle{\Box{7pt}}}

\vbadness=10000

\def\abstract#1{
\vskip.5in\vfil\centerline
{\bf Abstract}\penalty1000
{{\smallskip\ifx\answ\bigans\leftskip 2pc \rightskip 2pc
\else\leftskip 5pc \rightskip 5pc\fi
\noindent\abstractfont \baselineskip=12pt
{#1} \smallskip}}
\penalty-1000}
%
\lref\GKP{
  S.~S.~Gubser, I.~R.~Klebanov and A.~M.~Polyakov,
  ``Gauge theory correlators from non-critical string theory,''
  Phys.\ Lett.\ B {\bf 428}, 105 (1998)
  [hep-th/9802109].}
\lref\Wittenone{E.~Witten,
  ``Anti-de Sitter space and holography,''
  Adv.\ Theor.\ Math.\ Phys.\ {2} (1998) 253
  [hep-th/9802150].}
\lref\CJ{S.~Coleman and R.~Jackiw ,
``Why Dilatation Generators Do Not Generate Dilatations,''
  Ann.\ Phys.\ (N.Y.) {\bf 67},  (1971) 552.}
\lref\Coll{J.~C.~Collins , ``Renormalization of the energy-momentum tensor in
$\Phi^4$ theory,''
 Phys.\ Rev.\ D {\bf 14}, (1976) 1965.} 
\lref\KW{
  I.~R.~Klebanov and E.~Witten,
  ``AdS/CFT correspondence and symmetry breaking,''
  Nucl.\ Phys.\ B {\bf 556}, 89 (1999)
  hep-th/9905104].}
\lref\HR{
  T.~Hartman and L.~Rastelli,
  ``Double-trace deformations, mixed boundary conditions and functional
  determinants in AdS/CFT,''
  hep-th/0602106.}
\lref\BFS{
  M.~Bianchi, D.~Z.~Freedman and K.~Skenderis,
  ``How to go with an RG flow,''
  JHEP {\bf 0108}, 041 (2001) [hep-th/0105276];
  ``Holographic renormalization,''
  Nucl.\ Phys.\ B {\bf 631}, 159 (2002)
  [hep-th/0112119].}
\lref\Wittentwo{
  E.~Witten,
  ``Multi-trace operators, boundary conditions, and AdS/CFT correspondence,''
  arXiv:hep-th/0112258.}
\lref\HM{
  T.~Hertog and K.~Maeda,
  ``Black holes with scalar hair and asymptotics in N = 8 supergravity,''
  JHEP {\bf 0407}, 051 (2004)
  [hep-th/0404261].}
\lref\HMTZone{
  M.~Henneaux, C.~Martinez, R.~Troncoso and J.~Zanelli,
  ``Asymptotically anti-de Sitter spacetimes and scalar fields with a
  logarithmic branch,''
  Phys.\ Rev.\ D {\bf 70}, 044034 (2004)
  [hep-th/0404236].}
\lref\HMTZtwo{
  M.~Henneaux, C.~Martinez, R.~Troncoso and J.~Zanelli,
  ``Asymptotic behavior and Hamiltonian analysis of anti-de Sitter gravity
  coupled to scalar fields,''
  arXiv:hep-th/0603185.}
\lref\DS{
  S.~Deser and A.~Schwimmer,
  ``Geometric classification of conformal anomalies in arbitrary dimensions,''
  Phys.\ Lett.\ B {\bf 309}, 279 (1993)
  [hep-th/9302047].}
\lref\EGPR{
  S.~Elitzur, A.~Giveon, M.~Porrati and E.~Rabinovici,
  ``Multitrace deformations of vector and adjoint theories and their
  holographic duals,''
  JHEP {\bf 0602}, 006 (2006)
  [hep-th/0511061].}
\lref\CW{
  S.~R.~Coleman and E.~Weinberg,
  ``Radiative Corrections As The Origin Of Spontaneous Symmetry Breaking,''
  Phys.\ Rev.\ D {\bf 7}, 1888 (1973).}
\lref\ISTY{
  C.~Imbimbo, A.~Schwimmer, S.~Theisen and S.~Yankielowicz,
  ``Diffeomorphisms and holographic anomalies,''
  Class.\ Quant.\ Grav.\  {\bf 17}, 1129 (2000)
  [hep-th/9910267].}
\lref\STone{
  A.~Schwimmer and S.~Theisen,
  ``Diffeomorphisms, anomalies and the Fefferman-Graham ambiguity,''
  JHEP {\bf 0008}, 032 (2000)
  [hep-th/0008082].}
\lref\ABS{
  O.~Aharony, M.~Berkooz and E.~Silverstein,
  ``Multiple-trace operators and non-local string theories,''
  JHEP {\bf 0108}, 006 (2001)
  [hep-th/0105309];
  ``Non-local string theories on AdS(3) x S**3 and stable  non-supersymmetric
  backgrounds,''
  Phys.\ Rev.\ D {\bf 65}, 106007 (2002)
  [hep-th/0112178].}
\lref\BSS{
  M.~Berkooz, A.~Sever and A.~Shomer,
  ``Double-trace deformations, boundary conditions and spacetime
  singularities,''
  JHEP {\bf 0205}, 034 (2002)
  [hep-th/0112264].}
\lref\Muck{
  W.~Muck,
  ``An improved correspondence formula for AdS/CFT with multi-trace
  operators,''
  Phys.\ Lett.\ B {\bf 531}, 301 (2002)
  [hep-th/0201100].}
\lref\Minces{
  P.~Minces,
  ``Multi-trace operators and the generalized AdS/CFT prescription,''
  Phys.\ Rev.\ D {\bf 68}, 024027 (2003)
  [hep-th/0201172].}
\lref\Petkou{
  A.~C.~Petkou,
  ``Boundary multi-trace deformations and OPEs in AdS/CFT correspondence,''
  JHEP {\bf 0206}, 009 (2002)
  [hep-th/0201258].}
\lref\SSh{
  A.~Sever and A.~Shomer,
  ``A note on multi-trace deformations and AdS/CFT,''
  JHEP {\bf 0207}, 027 (2002)
  [hep-th/0203168].}
\lref\Barbon{
  J.~L.~F.~Barbon,
  ``Multitrace AdS/CFT and master field dynamics,''
  Phys.\ Lett.\ B {\bf 543}, 283 (2002)
  [hep-th/0206207].}
\lref\KP{
  I.~R.~Klebanov and A.~M.~Polyakov,
  ``AdS dual of the critical O(N) vector model,''
  Phys.\ Lett.\ B {\bf 550}, 213 (2002)
  [hep-th/0210114].}
\lref\BF{
  P.~Breitenlohner and D.~Z.~Freedman,
  ``Positive Energy In Anti-De Sitter Backgrounds And Gauged Extended
  Supergravity,'' Phys.\ Lett.\ B {\bf 115}, 197 (1982);
  ``Stability In Gauged Extended Supergravity,''
  Annals Phys.\  {\bf 144}, 249 (1982).}
\lref\FMMR{
  D.~Z.~Freedman, S.~D.~Mathur, A.~Matusis and L.~Rastelli,
  ``Correlation functions in the CFT($d$)/AdS($d+1$) correspondence,''
  Nucl.\ Phys.\ B {\bf 546}, 96 (1999)
  [hep-th/9804058].}
\lref\HS{
  M.~Henningson and K.~Skenderis,
  ``The holographic Weyl anomaly,''
  JHEP {\bf 9807}, 023 (1998)
  [hep-th/9806087].}
\lref\GK{
  S.~S.~Gubser and I.~R.~Klebanov,
  ``A universal result on central charges in the presence of double-trace
  deformations,''
  Nucl.\ Phys.\ B {\bf 656}, 23 (2003)
  [hep-th/0212138].}
\lref\GM{
  S.~S.~Gubser and I.~Mitra,
  ``Double-trace operators and one-loop vacuum energy in AdS/CFT,''
  Phys.\ Rev.\ D {\bf 67}, 064018 (2003)
  [hep-th/0210093].}
\lref\Zumino{
  B.~Zumino, 
  ``Effective Lagrangians And Broken Symmetries,''
  in `Lectures on Elementary Particles and Quantum Field Theory', 
  S. Deser, M. Grisaru, H. Pendleton (eds.), 
  M.I.T. Press 1970.}
%
 
\Title{\vbox{
\rightline{\vbox{\baselineskip12pt}}}}
{Remarks on Resonant Scalars in the }
\vskip-1cm
{\titlefont\centerline{AdS/CFT Correspondence
\footnote{$^{\scriptscriptstyle*}$}{\sevenrm
Partially supported by GIF,
the German-Israeli Foundation for Scientific Research, 
the Minerva Foundation, DIP, the German-Israeli Project Cooperation, 
by grants FONDECYT 1060648 and
7020832 (M.B.) and 
by the European Research and Training Networks 
`Superstrings' (MRTN-CT-2004-512194) (A.S.) and 
`Forces Universe' (MRTN-CT-2004-005104) (S.T.) 
}}}
\vskip 0.3cm
\centerline{M. Ba\~nados$^a$,  A.~Schwimmer$^b$ and
S.~Theisen$^c$ }
\vskip 0.6cm
\centerline{$^a$ \it Departamento de F\'\i sica, P. Universidad
Cat\'olica de Chile, Casilla 306, Santiago 22, Chile}
\vskip.2cm
\centerline{$^b$ \it Department of Physics of Complex Systems,
Weizmann Institute, Rehovot 76100, Israel}
\vskip.2cm
\centerline{$^c$ \it Max-Planck-Institut f\"ur Gravitationsphysik,
Albert-Einstein-Institut, 14476 Golm, Germany}
\vskip0.0cm

\abstract{
The special properties of scalars having a mass such that the two possible
dimensions of the dual scalar respect the unitarity and the 
Breitenlohner-Freedman bounds and their ratio is integral 
(``resonant scalars'')
are studied in the AdS/CFT correspondence.
The role of logarithmic branches in the  gravity theory is related to
the existence of a trace anomaly and to a marginal deformation
in the Conformal Field Theory. The existence of asymptotic
charges for the conformal group in the gravity
theory is interpreted in terms  of the properties of the corresponding CFT.}
\Date{\vbox{\hbox{\sl {April 2006}}
}}
\goodbreak

\newsec{Introduction}

The coupling of scalars to gravity plays an important role in the study of
the AdS/CFT correspondence. The scalar fields in gravity are related to 
scalar operators in the CFT and therefore  test its detailed  structure.
 Consider the general Lagrangian  which admits an AdS background solution
coupling gravity to a scalar field $\Phi$ in $D$ dimensions:
\eqn\action{
S=\int d^d x d\rho\sqrt{G}\left[R+\Lambda-G^{\mu\nu}\p_\mu\Phi\p_\nu\Phi
-m^2\Phi^2-V(\Phi)\right]}
where we separated the $D$ coordinates into $d=D-1$ ``boundary'' 
coordinates and the ``radial'' coordinate
$\rho$. For simplicity we did not include higher order curvature terms in the
gravity part. The potential of the scalar is separated into a ``mass term''
and an interaction term $V(\Phi)$ which will be specified later.
Neglecting the scalar interaction term the possible dimensions of the
dual operators in the CFT are 
are:
\eqn\dims{
\Delta_\pm={d\over2}\pm\nu}
where
$$
\nu=\sqrt{{d^2\over4}+m^2}\,.
$$
We will be interested in this note in the situation when both $\Delta_+$
and $\Delta_-$ represent normalizable modes in the gravity corresponding
to operators which satisfy the unitarity bound in the CFT. The condition
for this to happen is:
\eqn\condunitary{
0\leq\nu<1}
where the lower limit is the ``Breitenlohner-Freedman'' 
\BF\
bound.
As discussed in \Wittentwo,\KW,\HR,\Wittenone,\FMMR,\GK,\GM\
this situation corresponds to two possible CFT's in which 
only one of the operators ${\cal O}_{\pm}$ with 
dimensions $\Delta_{\pm}$, respectively, exists. A flow produced by
the relevant perturbation $f\int {\cal O}_-^2 d^dx$ 
takes the two theories into each other. Moreover, there is an 
interesting ``duality'' between 
the two theories, the generating functionals for correlators of the 
${\cal O}_\pm$ operators in the two theories being Legendre 
transforms of each other.

Special, very interesting features appear when the scalars are ``resonant''
i.e. if
\eqn\resonant{
{\Delta_+\over\Delta_-}=n}
where $n$ is an integer. In the gravity theory solutions with
logarithmic dependence on the
radial coordinate appear. Some implications 
for the AdS/CFT correspondence of the resonant scalars in the $n=1$
case  were discussed in \BFS,\Wittentwo.
Recently an in depth study of the gravity theory in this situation 
was undertaken in \HM,\HMTZone,\HMTZtwo. 
In particular the existence of asymptotic
charges of the conformal group was studied. In order that these charges 
would be well defined, certain  ``integrability conditions'' relating the 
coefficients of the expansion near the boundary should be imposed.
In the present note we study the general implications of ``resonant scalars''
for the AdS/CFT correspondence. When the resonance condition 
\resonant\ is fulfilled, if the operator ${\cal O}_-$ exists in the theory, 
necessarily also the operator ${\cal O}_+$ is present in the same theory 
since it appears in the 
OPE of $n$ ${\cal O}_-$ operators \foot{We limit ourselves to the discussion
of this setup, though in principle one could have a theory where  starting 
with the ${\cal O}_+$ operator the ${\cal O}_-$ operator is not present.}.
As a consequence of the presence of both operators,
two features single out this type CFT:

\noindent
a) there is a ``type B'' conformal anomaly involving one energy-momentum 
tensor and $n+1$ ${\cal O}_+$ operators;

\noindent
b) there is a marginal deformation generated by the interaction
\foot{The role of  marginal deformations in the 
AdS/CFT correspondence in d=4 and d=3 examples were studied by \Wittentwo\ and 
\EGPR.}
$f\int {\cal O}_-^{n+1} d^d x$. 

The ``running'' of the interaction b) is related to the anomaly a).
The gravity theory faithfully reproduces these features. In particular
the conditions for the existence of asymptotic conformal charges can be 
understood as a cancellation of the conformal anomaly a) by a non-polynomial
``Wess-Zumino-Green-Schwarz'' term or, equivalently, b), by the existence of a 
marginal perturbation which could be made ``truly marginal'' by the addition
of a non-polynomial term.

The paper is organized as follows:
In section 2 we discuss the structure of CFT having resonant scalar operators.
We analyze the general structure of the type B trace anomaly and its 
implications. We study the marginal perturbation appearing, we 
calculate its lowest order running and we relate it to the trace anomaly.
In section 3 we study the gravity theory in Fefferman-Graham gauge in 
dimensional regularization. We show the appearance of the trace anomaly 
and we discuss in detail the relation between the existence of asymptotic
conformal charges  and the trace anomaly and the running of 
multitrace deformations.
In section 4 we summarize our results and we discuss possible 
future applications.
\bigskip

\newsec{Trace anomalies and marginal perturbations in CFT 
with resonant scalars}

Let's consider a CFT in which both operators ${\cal O}_+$ and ${\cal O}_-$ are
present with dimension given by \dims\ with conditions \condunitary\ and 
\resonant\ fulfilled.
We start with presenting  the mechanism for a trace anomaly \CJ\ which
is ``type B'' following the classification of \DS.
Consider the correlator $G(x_1,...,x_{n+1})$ of $n+1$ operators ${\cal O}_+$ 
\eqn\correlator{
G(x_1,\dots,x_{n+1})=\langle0|T({\cal O}_+(x_1)\dots
{\cal O}_+(x_{n+1}))|0\rangle\,.}
Since ${\cal O}_+$ appears in the OPE of $n$ operators 
${\cal O}_-$ it can be written as:
\eqn\defOplus{
{\cal O}_+(x)~\equiv~\, :\!{\cal O}_-^n(x)\!:}
where the normal ordering sign is used symbolically in order to indicate that 
inside \defOplus\
there are no short distance singularities.
The correlator $G$ can now be evaluated using the two point function of 
${\cal O}_-$:
\eqn\twopoint{
\langle {\cal O}_-(x){\cal O}_-(y)\rangle={c\over|x-y|^{2\Delta_-}}\,.}
The expression obtained does not have a well defined Fourier transform: 
indeed, after removing an overall $\delta$-function for momentum conservation
and taking into account that $(n+1)\Delta_-=d$, the dimension is $0$ 
indicating a logarithmic ultraviolet divergence. A subtraction at a
scale $\mu$ is needed
and the correlator will contain a factor $\log(p^2/\mu^2)$ where 
$p^2$ is the overall scale of the external momenta. 
Clearly, the presence of a scale shows that dilation invariance is 
violated by the correlator \correlator\
or that, equivalently, when an energy-momentum tensor is inserted,
the Ward identity following from the tracelessness of the 
energy-momentum tensor is violated \CJ. 
\bigskip
The same information can be summarized by looking at the generating 
functional when the energy-momentum tensor of the theory is coupled 
to an external gravitational field $g_{ij}$ and the operator ${\cal O}_+$ 
to a source $\alpha$ of dimension $\Delta_-$:
\eqn\genfunct{
\exp[iW(g,\alpha)]=\int {\cal D}\varphi\, e^{i S_0(g,\varphi)
+i\int d^d x \sqrt{g}\alpha {\cal O}_+}}
where $\varphi$ represent the fields of the CFT, $S_0$ is the
action coupled to the external metric and $W$ is the generating functional.
Now the diagram  
\figinsert\figone{}{1.0in}{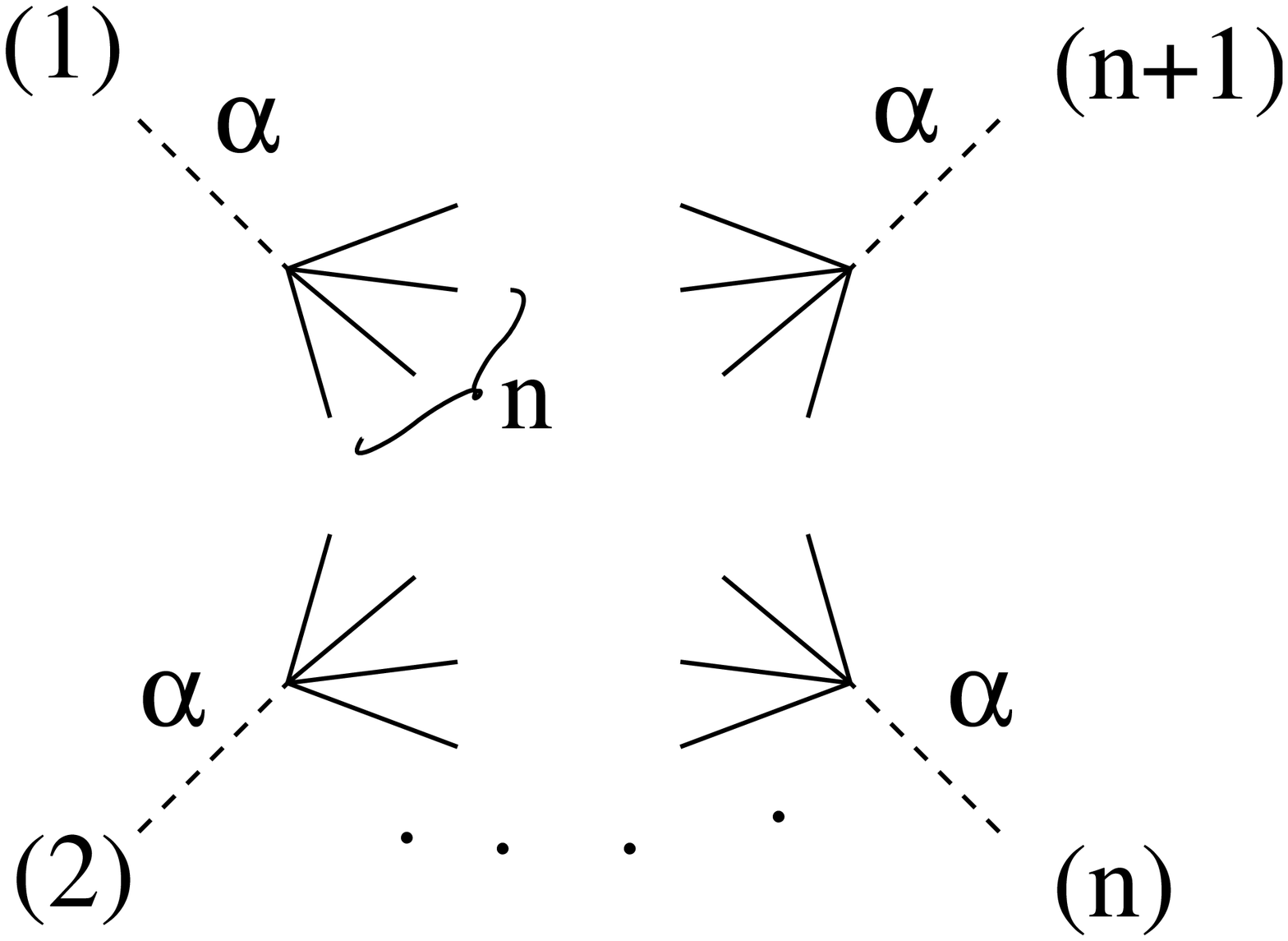}
\noindent
will contribute a term 
\eqn\genfuncta{
\int d^d x\, \alpha^p\log(\Laplace_0/\mu^2)\alpha^{n+1-p}}
to $W$ before coupling to $g$. $\Laplace_0$ is the Laplacian in flat 
space. Since  general coordinate invariance is 
preserved this term generates in the full $W$ a term
\eqn\genfunctb{
\int d^d x\sqrt{g}\,\alpha^p\log(\Laplace/\mu^2)\alpha^{n+1-p}}
where this time $\Laplace $ is the full, covariant Laplacian and additional,
higher order terms.
Performing a Weyl transformation on $W$:
\eqn\Weyl{\eqalign{
g_{ij}(x)&\to g_{ij}(x)e^{2\sigma(x)}\cr
\alpha(x)&\to\alpha(x)e^{-\Delta_-\sigma(x)}}}
we obtain for the variation of $W$, $\delta_{\sigma} W$:
\eqn\deltaW{
\delta_\sigma W=-2\int d^d x\sqrt{g}\,\sigma(x)\alpha^{n+1}(x)}
due to the transformation of the covariant Laplacian.

The expression \deltaW\ represents the type B trace anomaly in this case.
Though the effective action contains a scale $\mu$, indicating an explicit
breaking of scale invariance, the variation \deltaW\ itself is scale free 
and has the usual properties of an anomaly, i.e. it is a local variation
fulfilling the Wess-Zumino condition of a nonlocal effective action.
The Wess-Zumino condition is fulfilled in a way characteristic to 
all type B anomalies, i.e. the variation $\delta_\sigma W$ itself 
is Weyl invariant.
We stress that the presence of this anomaly does not spoil the 
conformal invariance of the theory: the energy-momentum tensor 
is traceless as a 
consequence of the Heisenberg equations of motion and the Ward 
identity following from the tracelessness of the energy-momentum tensor
is obeyed in all correlators except one class of anomalous ones
whose coefficients are, however, related by the Wess-Zumino condition. 
A special important feature of the type B anomaly is that due to the presence
of a scale in the effective action not only the Ward identity involving
the dilation current is violated but also the conservation of the 
dilation charge (unlike for type A trace anomalies, or axial anomalies,
for that matter). 

The anomaly \deltaW\ is the analogue of the type B trace anomaly involving
only energy-momentum tensors. An important difference is that \deltaW\ exists
in any integer dimension while the one involving only energy-momentum tensors 
exists only in even dimensions.

In dimensional regularization type B anomalies are signaled by 
the presence in the effective
action of terms having a factor ${1\over (d-p)}$ with 
nonzero residue where $p$ is an integer.

Though we will not use them in this note, we add for completeness 
a short discussion of the generalizations of \deltaW.
As mentioned above the anomaly satisfies the Wess-Zumino condition by being
Weyl invariant. Therefore a complete list of trace anomalies can be obtained
by construction of Weyl invariant, local expressions involving a scalar source
$\gamma$ with dimension $\Gamma$ \Zumino. 
If   ${d\over \Gamma}=n+1$
where $d$ is an integer dimension and $n$ is an integer the anomaly is given
by \deltaW\ with $\alpha$ replaced by $\gamma$. 
Let's consider the situation ${(d-2)\over\Gamma}=p+1$ with 
$d$ and $p$ integers, which would illustrate the general procedure.
Consider a ``composite''  metric  $\bar g_{ij}$ defined by    
\eqn\gbar{
\bar g_{ij}\,\equiv\,g_{ij}\gamma^{{2(p+1)\over(d-2)}}\,.}
By construction $\bar g_{ij}$ is invariant under a Weyl transformation and
therefore any expression constructed from it will be also Weyl invariant.
Consider therefore as a candidate for the anomaly 
\eqn\Sbar{
\int d^d x\,\sigma(x)\sqrt{\bar g}\bar R}
where $\bar R$ is the curvature scalar constructed from $ \bar g_{ij}$.
Expressing $\bar R $ as a function of 
$R$ and $g_{ij}$ we obtain finally for the anomaly:
\eqn\Anomaly{
\int d^d x\sigma(x)\sqrt{g}\left\lbrace\gamma^{p+1}R
+{2(p+1)(d-1)\over d-2}\gamma^p\Laplace\gamma
+{(p^2-1)(d-1)\over d-2}\gamma^{p-1}\p_j\gamma\p^j\gamma\right\rbrace\,.}
This particular anomaly is  generated by the non-leading logarithmic divergence
in the quadratically divergent correlator of $p+1$ operators whose source
is $\gamma$. Similarly, by considering quadratic expressions in the curvatures
we can construct the anomalies when the dimensions of the source fulfill
$\Gamma={d-4\over p+1}$ etc.

We discuss now the other characteristic feature of CFT having 
resonant scalars. For such a theory 
\eqn\Sint{
S_{\rm int}\,\equiv\,{f\over(n+1)!}\int d^d x\,:\!{\cal O}_-^{n+1}\!:}
represents a marginal perturbation. Again the definition of the operator
appearing is given by the OPE of $ n+1$ ${\cal O}_-$ operators. Due to the
associativity of the OPE the operator can be given alternatively as
\eqn\Oalt{
:\!{\cal O}_-{\cal O}_+\!:\,.}
Generically the coupling constant will be renormalized. If $n+1$ is
a prime number the lowest contribution is again given by Fig. 1
with ${\cal O}_-$ replacing the source $\alpha$. The calculation is 
identical with the one giving the trace anomaly and the renormalized
coupling $f_R$ is
\eqn\fR{
f_R=f+c f^{n+1}\log(\Lambda^2/\mu^2)}
where $f$ is the bare coupling, producing a $\beta$ function:
\eqn\betafunction{
\beta(f_R)=c f_R^{n+1}\,.}
This relation between the trace anomaly and the renormalization 
of the coupling is a manifestation of the well known Ward identity \Coll
\eqn\Tjj{
T_j{}^j=\beta(f_R){\cal O}_-^{n+1}\,.}
A convenient way to define the renormalized coupling $\bar f$ is through
the effective potential of ${\cal O}_-$ \CW,
$V_{\rm eff}(O_-)$:
\eqn\fbar{
\bar f\,\equiv\,\mu^{-d}V_{\rm eff}({\cal O}_-)
\Big|_{{\cal O}_-=\mu^{\Delta_-}}\,.}
Then following Coleman-Weinberg \CW\ the Callan-Symanzik equation for the
effective potential can be integrated. We use the $\beta$ function given
by \betafunction\ and we don't include a possible anomalous dimension
for ${\cal O}_-$ which would enter at higher order. The solution
of the Callan-Symanzik equation gives the form of the effective potential
to the order we are working:
\eqn\effpot{
V_{\rm eff}({\cal O}_-)=\bar f{\cal O}_-^{n+1}
+{c\bar f^{n+1}\over\Delta_-}{\cal O}_-^{n+1}
\log\left({{\cal O}_-\over \mu^{\Delta_-}}\right)}
which we will use in the comparison with the gravity treatment. 

\newsec{Resonant scalars in the gravity theory}
The results of the analysis described in this section agree 
with the ones described in \HMTZtwo.

We consider a gravity theory which admits an AdS solution 
coupled to a scalar $\Phi$  with the action \action.
We will analyze the solutions of \action\ in the Fefferman-Graham gauge
where the metric has the form:
\eqn\FGgauge{
ds^2={(d\rho)^2\over 4\rho^2}+{1\over\rho}g_{ij}(x,\rho)dx^i dx^j\,.}
This gauge is suited for the studies of anomalies signaled in dimensional
regularization by the appearance of poles in $d-N$ where $N$ is an integer,
in the solutions for various fields and the action evaluated on solutions
\ISTY,\STone,\HS. Since the integration measure contains 
$\rho^{-1-d/2}$, the integration can produce poles if the power $\rho^{N/2}$
appears in the expansion of the action. This condition
will be guiding us in setting up the regularization. By requiring
that we are in the resonant scalar situation (i.e. equations 
\dims,\condunitary\ and \resonant\
are fulfilled)  the expansion of the fields will have the general form:
\eqn\FGexpansions{\eqalign{
g_{ij}(x,\rho)&=\gz_{\!\!ij}(x)+\dots+
\gN_{\!\!\!\!\!ij}(x)\rho^{N/2}+\dots\cr
\Phi(x,\rho)&=\alpha(x)\rho^\Delta+\dots+\beta(x)\rho^{n\Delta}+\dots}}
where the dots include additional terms which should be present due 
to the coupled  equations of motion (like $\rho^{2\Delta}$, 
possibly half-integer powers, etc.) which do not play,
however, a role in the anomaly structure we are studying. 
In \FGexpansions\ and from now on we denote:
\eqn\defDelta{
\Delta\equiv{\Delta_-\over 2}\,.}
In order that the critical $\rho$ dependence appears we require:
\eqn\requirement{
(n+1)\Delta={N\over 2}\,.}
 The regularization choice \requirement\ violates \resonant\ by a term of
order $d-N$ and therefore it is recovered only in the limit.

The linear part in the $\Phi$ equation of motion fixes then $m$ through
eq.\dims\ to be:
%
\eqn\mass{m^2={N^2\over(n+1)^2}\left[1-{(n+1)d\over N}\right]\,.}
Finally we choose $V(\Phi)$ in eq.\action\ such that the second term in the 
expansion of $\Phi$ in \FGexpansions\ 
appears in leading order in the nonlinearity:
\eqn\pot{V(\Phi)={h\over n+1}\Phi^{n+1}\,.}
Generically the potential may have additional terms which do not 
change the anomaly structure. 

The procedure is now straightforward:
We study the coupled equations of motion following from the action \action\
with the choice \mass\ for the expansions \FGexpansions; 
once the solutions are 
known we use them in the action isolating the terms which have poles 
in $d-N$ . We defer a more general discussion for arbitrary actions involving
$g_{ij}$ and $\Phi$ based on diffeomorphisms in $d+1$ dimensions to another
publication.

The case $n=1$ when ${\cal O}_+$ and ${\cal O}_-$ coincide requires a special 
treatment which will be given after the general discussion.

Now, the $\Phi$ equation of motion is:
\eqn\eomPhi{
\Laplace\Phi=m^2\Phi+{h\over2}\Phi^n\,.}
The linear term determines the relation between $m$ and 
$\Delta$ which, due to our condition \requirement, gives \mass.
From the leading nonlinear matching we obtain:
\eqn\solbeta{
\beta(x)={h(n+1)\over 2N(n-1)}{\alpha(x)^n\over N-d}+\beta_0(x)\,.}
The appearance of the pole in $d-N$ signals the existence of the
logarithmic branch. The function $\beta_0$  is undetermined by the expansion
and it is the analogue of the so called ``Fefferman-Graham ambiguity''
in this context.

The  gravitational equation of motion is:
\eqn\eomgrav{
R_{\mu\nu}-{1\over2}G_{\mu\nu}R-{\Lambda\over2}G_{\mu\nu}
-\p_\mu\Phi\p_\nu\Phi+{1\over2}G_{\mu\nu}\left(
\p_\alpha\Phi\p^\alpha\Phi+m^2\Phi^2+{h\over n+1}\Phi^{n+1}\right)=0\,.}
Instead of using it directly we will concentrate on the terms of the action
which get contributions from the scalar field. First, using the trace of
eq. \eomgrav\ we can express $R$ through the other terms and the 
action becomes:
\eqn\Sinsert{S=\int d^d x d\rho\,\rho^{-1-d/2}\sqrt{g}
\left[ d+{1\over d-1}\left(m^2\Phi^2+{h\over n+1}\Phi^{n+1}\right)\right]}
where we used the relation between $\Lambda$ and the AdS radius in the 
parametrization  \FGgauge:
\eqn\defLambda{
\Lambda=-d(d-1)\,.}
Now, in \Sinsert\ there are two contributions depending on the scalar field:
the one appearing explicitly and the other in back-reaction to $\sqrt{g}$.
The interesting term in the back-reaction after expanding $\sqrt{g}$
around $\gz$ is contributed by 
${\displaystyle{1\over2}\gz_{\!\!ij}\gN_{\!\!\!\!ij}}$
and can be calculated through its relation to $R$ where $\delta R$ is 
the part of the coefficient of $\rho ^{N/2}$ in the expansion of $R$
depending on $\Phi$ :
\eqn\deltaR{
{1\over2}\tr\big(\!\!\gN\!\!\big)={\delta R\over 2N(N-d-1)}\,.}
Putting the two contributions together we have to 
isolate the coefficient of $\rho^{N/2}$ in the expression
\eqn\coeffrho{
{d\over2N(N-d-1)}\p_\alpha\Phi\p^\alpha\Phi
+{1\over d-1}\left[1+{d(d+1)\over 2N(N-d-1)}\right]
\left(m^2\Phi^2+{h\over n+1}\Phi^{n+1}\right)\,.}
The result is:
\eqn\result{\eqalign{
&{(N-d)\over(N-d-1)(d-1)(n+1)^2}
\Big[ d^2(n+1)+2 N(N-1)-d(n+1)(2N-1)\Big]\alpha\beta\cr
\noalign{\vskip.2cm}
&\qquad\qquad\qquad\qquad
+{d(d+1-2N)+2N(N-1)\over 2N(N-d-1)(d-1)}{h\over n+1}\alpha^{n+1}\,.}}
We remark the $(d-N)$ factor in the $\alpha \beta$ term which shows that
even though there is a pole term in the expression \solbeta\ of 
$\beta$ in terms of $\alpha$  the integrand of $S$ finally does not have 
any explicit pole terms. The pole term is the result of the integration
over $\rho$. Keeping the pole and finite term the structure of the result
is:
\eqn\Sab{
S[\gz_{ij},\alpha,\beta_0]=\int\sqrt{g^{(0)}}d^d x
\left[{c_1\over d-N}h \alpha^{n+1}+c_2\a\beta_0\right]}
where $c_{1,2}$ are numerical coefficients.

We discuss now briefly 
the exceptional case $n=1$ when $\Delta=N/4$.
In this case the expansion \FGexpansions\ is replaced by 
\eqn\Philog{
\Phi(x,\rho)=\left[{\alpha(x)\over d-N}+\beta_0(x)\right]\rho^{N/4}}
and it is easy to verify, following the steps outlined above, that again
\Sab\ is obtained  with $n=1$ \BFS.
 
Now the meaning of \Sab\ is obvious \ISTY,\STone:
the presence of the pole requires a 
subtraction before the limit $d\to N$ is taken and as a consequence
the subtracted action is nonlocal and it is no longer Weyl invariant.
The Weyl variation
can be calculated before taking the limit and it is finite and nonvanishing:
\eqn\deltaS{\delta_\sigma S[\gz,\a,\beta_0]
=-c_1 h\int d^d x\sqrt{g^{(0)}}\sigma\alpha^{n+1}
=\delta_\sigma\left\lbrace{c_1\over2}h\int d^d x\sqrt{g^{(0)}}
\alpha^{n+1-p}\log(\Laplace/\mu^2)\alpha^p\right\rbrace\,.}
The structure of the basic result \Sab\ agrees completely with the structure
of the CFT discussed in section 2  where $\alpha$ in the gravity
theory should be identified with the source of the ${\cal O}_+$ operator
in the CFT. In particular the trace anomaly \deltaW\
is reproduced by \deltaS. Therefore we conclude that also in the 
case of resonant scalars there is a relation between two theories:
a CFT with operators ${\cal O}_-$ and ${\cal O}_+$ 
and a gravity theory with action \action\
and mass parameter such that the conditions \dims,\condunitary\ and \resonant\
are satisfied.  The exact duality between the two theories 
requires a matching of the numerical coefficients in the above formulae
which would depend on the exact formulation of the CFT and 
more specification of the gravity background 
including the additional compactified dimensions, etc.
We stress that both theories are well defined and conformal, the logarithmic
branches signaling, as we discussed in detail above, 
the matched conformal anomalies.
 
We discuss now the interesting observation made in 
\HM,\HMTZone,\HMTZtwo\ that in the gravity
theory the asymptotic conformal charges are well defined only if a certain
integrability condition is obeyed expressing the function $\beta_0$  
appearing in \Sab\ in terms of $\alpha$.
The integrability condition in our notation is the condition that 
the total Weyl variation of \Sab\ vanishes:
\eqn\deltaSS{
\delta_\sigma S=\int d^d x\sqrt{g^{(0)}}\Big\lbrace
\sigma(x)c_1 h\alpha^{n+1}+\alpha c_2\delta_\sigma\beta_0
+\alpha n \Delta_-c_2\sigma(x)\beta_0\Big\rbrace=0}
or, more explicitly:
\eqn\deltab{
c_2\delta_\sigma\beta_0=-c_1\sigma(x)h\alpha^n(x)
-n\Delta_- c_2\sigma(x)\beta_0\,.}
A particular solution of \deltab\ proposed in \HM,\HMTZone,\HMTZtwo\ is
\eqn\solb{
\beta_0={c_1\over c_2\Delta_-}h\alpha^n(x)
\log\left({\alpha(x)\over\mu^{\Delta_-}}\right)}
where the choice of the scale $\mu$ includes the ambiguity of adding
to \solb\ a term  $\alpha^n$ with an arbitrary coefficient.
From the point of view of the gravity this condition is quite natural:
the existence of well defined asymptotic charges implies that
under the solutions which satisfy the requirement the action is 
invariant on the full conformal group. Of course, once \deltab\ is satisfied
the two terms in \Sab, after the subtraction,
combine, the scale $\mu$ disappears from the expression and $S$ becomes:
\eqn\SS{
S=\int d^d x\sqrt{g^{(0)}}\left\lbrace c_1 h\alpha^{n+1-p}
\log\left({\Laplace\over\alpha^{2/\Delta_-}}\right)\alpha^p
+\bar c_1\alpha^{n+1}\right\rbrace}
which is Weyl invariant. In  \SS\ we made explicit the
freedom of adding a conformally invariant term as a consequence of 
the ambiguity in \solb.

From the point of view of the CFT dual to the gravity theory the
interpretation is obvious: 
if  the generating functional $W$ is identified with $S$ as given by \SS\
a term was added to the anomalous, one loop term \genfuncta. This term which 
originates in the $\alpha\beta_0$ expression with the choice \solb\
for $\beta_0$ is classical and it is tuned such as to cancel the anomaly.
Therefore it is the conformal version of a Green-Schwarz-Wess-Zumino term.
Of course the term added, though local, is non-polynomial so in a usual
QFT situation we would not consider it legal.

An alternative, closely related interpretation of the integrability condition
is offered by looking at the solution 
\FGexpansions,\solbeta,\solb\ as representing
a perturbation of the original CFT by a ``multiple trace deformation''
\Wittentwo,\KW,\HR,\ABS,\BSS,\Muck,\Minces,\Petkou,\SSh,\Barbon,\KP.
The deformation is given by the integral of the
defining relation of $\beta_0$ :
\eqn\Spert{
S_{\rm pert}=\int d^d x\left\lbrace {h c_1\over c_2\Delta_-}
\left[{{\cal O}_-^{n+1}\over n+1}\log{{\cal O}_-\over\mu^{\Delta_-}}
-{{\cal O}_-^{n+1}\over (n+1)^2}\right]
+{f\over (n+1)!}{\cal O}_-^{n+1}\right\rbrace}
This deformation has two pieces: a marginal perturbation with arbitrary
strength which has a running
as we discussed in section 2 and the term containing the logarithm
of ${\cal O}_-$ which violates conformal invariance.
The first term will produce an  effective 
potential at lowest order of the form \effpot. Treating the second term of
\Spert\ as classical and choosing appropriately the bare coupling
and the scale the two logarithmic terms could cancel exactly making the 
perturbation truly marginal at lowest order. Therefore, at least at 
lowest order, we defined a new CFT which is a deformation of the original one.
To this new CFT corresponds the same conformally invariant action 
\SS\ possibly with a new interpretation. Indeed, we expect that due
to the perturbation the dimensions of the operators ${\cal O}_+$ 
and ${\cal O}_-$
will change such that they will not be anymore resonant and as a 
consequence the trace anomaly will no longer be present. Correspondingly
the interpretation of \SS\ will change which now represents the full generating
functional of the theory. The exact correspondence between
the operators in the perturbed CFT and $\alpha$ in \SS\ is an 
unsolved problem.

Summarizing, we offer two possible holographic interpretations of the 
requirement of well defined conformal charges in the gravity theory:
a) the original CFT with a Green-Schwarz-Wess-Zumino term added to the
partition function in order to cancel the trace anomaly and 
b) a perturbation of the original CFT by a marginal operator made truly 
marginal by canceling the one loop running of the perturbation by a fine tuned
term treated classically.
Given the connection between the trace anomaly and the running 
of the marginal perturbation discussed in  section 2 the two interpretation 
are closely related. 
 
\newsec{Discussion}

The presence of resonant scalars shows new features of the AdS/CFT 
correspondence. The two possible CFT's corresponding to the same gravity theory
with different boundary conditions are replaced by a single CFT where both
scalar operators are present. As a consequence the CFT has a type
B trace anomaly and related to it a marginal perturbation.
The gravity theory accommodates naturally this structure: 
the trace anomaly is reproduced and the boundary conditions allow the 
presence of the marginal perturbation through the ``multiple trace'' mechanism.
The action of the gravity theory corresponding to the situation 
when the asymptotic conformal charges are well defined and conserved 
corresponds to two possible CFT: one with the addition of a classical term
tuned to compensate the trace anomaly, the other a perturbation of the first
by a marginal operator made truly marginal by the addition
of another term treated classically.

The purpose of this note was to discuss the general structure of the AdS/CFT
correspondence in the above set up. We intend to study further this 
set up in concrete realizations.
At the level of the general structure several problems remain 
open which require further study:

\noindent
a) the calculation of correlators for both operators on 
the gravity side and their consistency;

\noindent
b) the validity of building  truly marginal perturbations of CFT 
through cancellation with terms treated classically beyond one loop and
the general consistency of such an approach;

\noindent
c) the detailed relation between the observables of a CFT built along the
lines mentioned in b) and the gravity action;

\noindent
d) the general structure of theories with scalars and higher order 
curvature terms following form their $D$ dimensional 
diffeomorphism invariance; 

\noindent
e) the appearance of scalar operators which obey 
the Breitenlohner-Freedman bound and produce higher trace anomalies 
of the type discussed in section 2.

\bigskip
\noindent
{\bf Acknowledgments:}
Very useful discussions with O. Aharony, M. Berkooz and M. Henneaux are
gratefully acknowledged.
AS thanks the Humboldt Foundation for its support and the 
Albert Einstein Institute for its hospitality.
ST thanks the Einstein Center and  the Weizmann Institute for 
their hospitality and support;

\listrefs

\bye